\documentclass[twocolumn,showpacs,showkeys,pre]{revtex4-1}
\usepackage{graphicx}
\usepackage{amsmath}
\usepackage{amsfonts}
\usepackage{amssymb}
\usepackage{multirow}
\usepackage{color}
\usepackage{comment}
\usepackage{physics}
\usepackage{mathtools}
\usepackage{epstopdf}

\begin{document}
\title{Extinction dynamics from meta-stable coexistences in an evolutionary game}
\author{Hye Jin Park}
\author{Arne Traulsen}
\affiliation{
Department of Evolutionary Theory, Max Planck Institute for Evolutionary Biology, 24306 Pl\"{o}n, Germany}

\begin{abstract}
Deterministic evolutionary game dynamics can lead to stable coexistences of different types.
Stochasticity, however, drives the loss of such coexistences. 
This extinction is usually accompanied by population size fluctuations.
We investigate the most probable extinction trajectory under such fluctuations by mapping a stochastic evolutionary model to a problem of classical mechanics using the Wentzel-Kramers-Brillouin (WKB) approximation. 
Our results show that more abundant types in a coexistence can be more likely to go extinct first well agreed with previous results, 
and also the distance between the coexistence and extinction point is not a good predictor of extinction.
Instead, the WKB method correctly predicts the type going extinct first. \end{abstract}

\maketitle
Evolutionary game dynamics describes the evolution of phenotypical traits~\cite{maynard-smith:book:1995a,nowak:Science:2004}. 
Evolution is based on birth and death processes, which are 
most adequately described by stochastic models~\cite{doebeli:eLife:2017}.
Most such models study a fixed population size \cite{moran:book:1962,nowak:Nature:2004,taylor:BMB:2004,traulsen:PRL:2005,assaf:PRL:2013}. 
An important concern about biological systems is the loss of types~\cite{Schodelbauerova:BC:2010,brook:nature:2000}. 
Consequently, persistence of phenotypes and probability
that an invader takes over a population have been explored. 
However, there is still a lack of understanding of the effect of population size fluctuations on extinction dynamics.

There is increasing evidence that fluctuations in population sizes 
caused by ecological processes can also affect evolutionary dynamics~\cite{pagie:JTB:1999,aviles:EER:1999,yoshida:Nature:2003,hauert:JTB:2006a,jones:AmNat:2009,cremer:PRE:2011,post:PTRSB:2009,hanski:PNAS:2011,sanchez:PLoSB:2013,papkou:Zoology:2016}.
Accordingly, understanding the effect of population size fluctuations becomes more important. 
Stochastic models as reaction systems with accompanying population size fluctuations have a long tradition in mathematical biology
\cite{murray:book:2007}.
Recently, a stochastic model~\cite{huang:PNAS:2015,czuppon:arXiv:2017} has been proposed, which directly connects evolutionary game dynamics to reaction rules by designing rates of death from interaction as a function of game payoffs. We examine persistence of types in this system under
the influence of population size changes. 
When population size is bounded, the population eventually goes to extinction due to stochasticity.
Before extinction of the whole population, the population looses individual types.
We examine single-type extinction from the coexistence of two types. 

Though stochastic noise can be approximated by white noise, the Fokker-Planck (FP) approach normally fails to
capture extinction properties because FP is not valid for large fluctuation~\cite{kessler:JSP:2007}.
To capture the extinction portrait in reaction systems, we use the method developed in~\cite{doi:JPA:1976,elgart:PRE:2004,kamenev:PRE:2008, kessler:JSP:2007}.
With this powerful tool, extinction dynamics for various systems has been described~\cite{ovaskainen:TREE:2010,elgart:PRE:2004, assaf:PRE:2006,kamenev:PRE:2008,meerson:PRE:2009,khasin:PRE:2010,lohmar:PRE:2011,gottesman:PRE:2012,gabel:PRE:2013,smith:PRE:2016,lohmar:jpam:2017}.
This method reformulates a master equation into a Hamilton-Jacobi equation form by using the Wentzel-Kramers-Brillouin (WKB) approximation.
An effective Hamiltonian which arises from the reformulated equation yields equations of motion and eventually the most probable extinction trajectory and its properties.

Typically in our model, one type is more likely to go extinct first, while the other becomes more abundant.
Interestingly, in some cases, a more abundant type in the coexistence is more likely to go extinct first, consistent with known results for the extinction dynamics in two-type populations \cite{gabel:PRE:2013}.
We use the same birth process, but a slightly different death process compared to~\cite{gabel:PRE:2013}.
The main differences are the death process from competition and the interpretation of its rate.
We interpret these terms as arising from game interactions, 
which are naturally connected to the competitive Lokta-Volterra dynamics in the deterministic limit~\cite{zeeman:PAMS:1995,hofbauer:book:1998}. 
For our system, we show that the distance from the coexistence to the extinction point is a better predictor for the type going extinct first than the abundances in the coexistence. 
Albeit the distance seems most important factor for the first extinction type, only WKB method gives the right answer for the first extinction type.

Following \cite{huang:PNAS:2015}, 
we consider three processes which trigger population size changes:
reproduction, spontaneous death, and death from competition. 
We focus on two types, $X$ and $Y$. 
The reproduction process can be described by the reactions,
	\begin{equation}
	\begin{aligned}
		&X \rightarrow X + X , \qquad
    		Y \rightarrow Y + Y ,
	\label{e.rep}
	\end{aligned}
	\end{equation}
with corresponding constant rate $\lambda_b$. 
Individuals die at a constant rate $\lambda_d~(<\lambda_b)$,
	\begin{equation}
	\begin{aligned}
		&X\rightarrow 0 , \qquad
		Y\rightarrow 0 .
	\end{aligned}
	\end{equation}
Due to the limitation of resources, individuals compete with each other.
There are four such reactions resulting in the death of one individual,
	\begin{equation}
	\begin{aligned}
	    &X+ X\rightarrow X, \qquad
	    Y+ Y\rightarrow Y,  \\
	   &  X+ Y\rightarrow X, \qquad
	    X+ Y\rightarrow Y,
	\label{e.contact}
	\end{aligned}
	\end{equation}
where the corresponding rates are determined by interactions between individuals. 
Inspired by evolutionary games, where outcome of interaction between individuals is represented by the game payoff matrix $A$, 
	\begin{equation}A=
	\begin{pmatrix}
	a & b\\
	c & d
	\end{pmatrix},
	\label{e.matrix}
	\end{equation}
we assume four positive
rate parameters, $a,b,c,$ and $d$,
	\begin{equation}
	\begin{aligned}
		&\lambda_{xx\rightarrow x} =\tfrac{1}{aM}\text{  ,  }\qquad
		\lambda_{yy\rightarrow y} =\tfrac{1}{dM}\text{  ,  }\\
		&\lambda_{xy\rightarrow x} =\tfrac{1}{cM}\text{  ,  } \qquad
		\lambda_{xy\rightarrow y} =\tfrac{1}{bM}
	\label{e.compete}.
	\end{aligned}
	\end{equation}
The element $A_{ij}$ of the payoff matrix means a payoff of the type $i$ from game interaction with an opponent type $j$.
Individual with a smaller payoff dies with higher probability in a direct competition \cite{huang:PNAS:2015}.
Parameter $M$ controls total population size in the quasi-steady state.
Since the chance that one individual meets another individual is proportional to population size $N$, 
and that competition rates are proportional to $1/M$, competition occurs at a rate $\mathcal{O}(N/M)$.
On the other hand, reproduction and spontaneous death occur in $\mathcal{O}(1)$.
If population size $N$ is much smaller than $M$, $N \ll M$, 
competition is negligible, and thus the population grows at a constant rate, $\lambda= \lambda_b-\lambda_d$.
For $N \gg M$, competition dominates other reactions, and the population size decreases until $N$ becomes comparable to $M$.
Therefore, population size $N$ is typically of the order of $M$. 

\begin{figure}[ht] 
\includegraphics[width=0.5\textwidth]{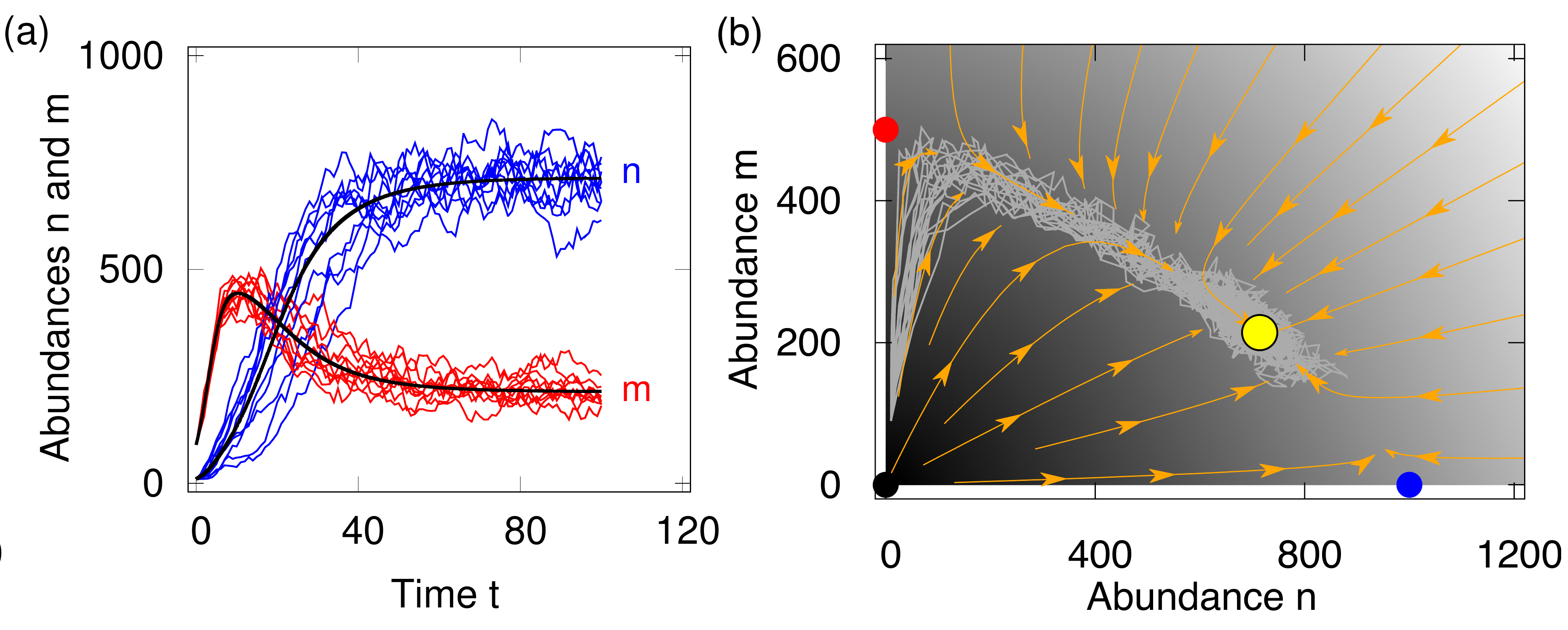}
\vspace{-0.5cm}
\caption{
(Color online) 
(a) Abundances versus time. 
The smooth thick lines are calculated from Eq.~(\ref{e.LV}), while red and blue lines are the stochastic simulation results.
Initially, abundances start from $n_0=10$ and $m_0=90$, and fluctuates around the stable fixed points.
(b) Abundance dynamics in $n$ and $m$.
Results in (a) are represented as grey lines.
Arrows show the direction of changing abundances, and background color indicates magnitude of change (darker color represents faster changes).
Dots are the fixed points of Eq.~(\ref{e.LV})
(parameters $M=2000$, $a=1$, $b=0.75$, $c=1.25$, $d=0.5$, $\lambda_b=0.6$, and $\lambda_d=0.1$).
}
\label{f.model}
\end{figure}

The system is described by abundances, $m$ and $n$, of types $X$ and $Y$.
For large populations, the reaction system can be described by deterministic equations~\cite{huang:PNAS:2015},
	\begin{equation}
	\begin{aligned}
		\dot{n} =  n \left(\lambda-\tfrac{n}{aM}-\tfrac{m}{bM}\right),\\
		\dot{m} =  m \left(\lambda-\tfrac{n}{cM}-\tfrac{m}{dM}\right).
	\end{aligned}
	\label{e.LV}
	\end{equation}
There are four fixed points, 
$\left(0, 0\right), \left(0, d\lambda M\right), \left(a\lambda M, 0\right),$ and $\frac{\lambda M}{bc-ad}\left[ac(b-d),bd(c-a)\right]$.  
The fixed point $(0,0)$ is unstable, and two fixed points on the axes are saddles.
For $a<c$ and $d<b$, the fixed point where $X$ and $Y$ coexist is stable (see Fig~\ref{f.model}) in the deterministic model. 

In the stochastic model, all reactions occur with constant rates with a Poisson process, and thus Poisson noise arises
due to the discrete nature of the number of individuals.
Hence, abundances fluctuate around the coexistence point, see Fig.~\ref{f.model},
until eventually one type goes extinct.  
Discrepancy between deterministic dynamics and stochastic dynamics arises from a process sequence.
We are interested in the most probable extinction trajectory 
starting from a coexistence state that is stable in the deterministic case ($a<c$ and $d<b$).
Stochastic systems are described by the probability $P_{n,m}(t)$ where the system is in state $(n,m)$ at a given time $t$.
The probability $P_{n,m}(t)$ changes according to the master equation
	\begin{equation}
	\begin{split}
	\frac{d{P}_{n,m}}{dt} 
	=&\lambda_b[(n\!-\!1)P_{n-1,m} + (m\!-\!1)P_{n,m-1} - (n\!+\!m)P_{n,m}]   \\
	+&\lambda_d[(n\!+\!1)P_{n+1,m} + (m\!+\!1)P_{n,m+1} - (n\!+\!m)P_{n,m}]  \\
	+&\tfrac{1}{aM}[(n+1)nP_{n+1,m}-n(n-1)P_{n,m}]  \\
	+&\tfrac{1}{bM}[(n+1)mP_{n+1,m}-nmP_{n,m}] \\
	+&\tfrac{1}{cM}[n(m+1)P_{n,m+1}-nmP_{n,m}] \\
	+&\tfrac{1}{dM}[(m+1)mP_{n,m+1}\!-\!m(m\!-\!1)P_{n,m}]  \\ 
	=&	\hat{H}P_{n,m}, 
	\end{split}\raisetag{4\baselineskip}
	\label{e.master}
	\end{equation}
where the effective Hamiltonian operator $\Hat{H}$ can be expressed by ladder operators for $P_{n,m}$ 
($\hat{a}^{\pm}P_{n,m}=P_{n\pm1,m}$ and $\hat{b}^{\pm}P_{n,m}=P_{n,m\pm1}$).
The probabilities become zero for all negative indices.
Note that $P_{0,0}(\infty)=1$, because extinction of both types is the final absorbing state in the stochastic model with a bounded population size.
The initial distribution quickly converges to the quasi-steady state which peaked at the coexistence point.
Subsequently, the probability leaks slowly into an absorbing state. 
Moreover, the extinction from the coexistence to a single-type population occurs much faster than 
the collapse of the whole population. 
Since we focus on $1\ll t \ll t_e$ where $t_e$ is a characteristic time for the collapse of a whole population, 
the probability leakage from the coexistence can be expressed with the characteristic time scale $\tau$ 
	\begin{equation}
	P_{n,m} = e^{-t/\tau} \psi_{n,m}~ \text{for}~n,m>0, 
	\label{e.prob}
	\end{equation} 
where $\psi_{n,m}$ is an eigenstate of $\hat{H}$ with eigenvalue $-1/\tau$,
corresponding to the quasi-stationary distribution.

Next, we obtain an effective Hamiltonian using the WKB method.
We start from the Eikonal ansatz with leading order
	\begin{equation}
	\psi_{n,m}=e^{-M S(x,y)},
	\label{e.eikonal}
	\end{equation}
where $S$ is a smooth function of relative abundances $x= n/M$ and $y= m/M$~\cite{gottesman:PRE:2012, gabel:PRE:2013, smith:PRE:2016}.
For large $M$, the Taylor expansion $S(x\pm \tfrac{1}{M}, y) \approx S(x,y)\pm\frac{1}{M} \partial_x S$ gives 
	\begin{equation}
	\begin{aligned}
	\hat{a}^\pm P_{n,m}&=P_{n\pm1,m} =e^{-t/\tau}e^{-M S(x\pm\tfrac{1}{M},y)}
	\approx P_{n,m} e^{\mp \partial_x S}.
	\label{e.taylor}
	\end{aligned}
	\end{equation}
Inserting the Eikonal ansatz~Eq.~(\ref{e.eikonal}) into Eq.~(\ref{e.master}) we obtain 
in the leading order for large $M$
	\begin{equation}
	1/(\tau M) + H(x, y, p_x, p_y)  = 0,
	\label{e.hj}
	\end{equation}
	%
with the effective Hamiltonian $H$ given by 
	\begin{equation}
	\begin{aligned}
		H=&\lambda_b[x(e^{p_x}-1)+y(e^{p_y}-1)]\\
		&+\lambda_d[x(e^{-p_x}-1)+y(e^{-p_y}-1)]\\
		&+\tfrac{1}{a}[x^2(e^{-p_x}-1)]+\tfrac{1}{b}[xy(e^{-p_x}-1)]  \\
		&+\tfrac{1}{c}[xy(e^{-p_y}-1)] +\tfrac{1}{d}[y^2(e^{-p_y}-1)],
	\end{aligned}
	\end{equation}
where $p_x = \partial_xS$ and $p_y = \partial_yS$.
It may seem surprising that the reaction system can be mapped into a Hamiltonian system.
In fact, the situation which is described by the master Eq.~(\ref{e.master}) can be interpreted as a particle in a potential well with noise.
As shown in Fig.~\ref{f.model}~(b), the speed of changing abundances depends on $n$ and $m$. 
If the abundances change fast, we can interpret this as the existence of a large potential gradient.
Fast changes of abundances give short waiting time, implying small $P_{n,m}$.
As a result, we can imagine a potential well with a minimum at the coexistence point. 
This potential landscape captures features of the probability $P_{n,m}$.
Therefore, momenta are related to the gradient of probabilities $P_{n,m}$~\cite{kessler:JSP:2007}.

\begin{figure}[t]
\includegraphics[width=0.5\textwidth]{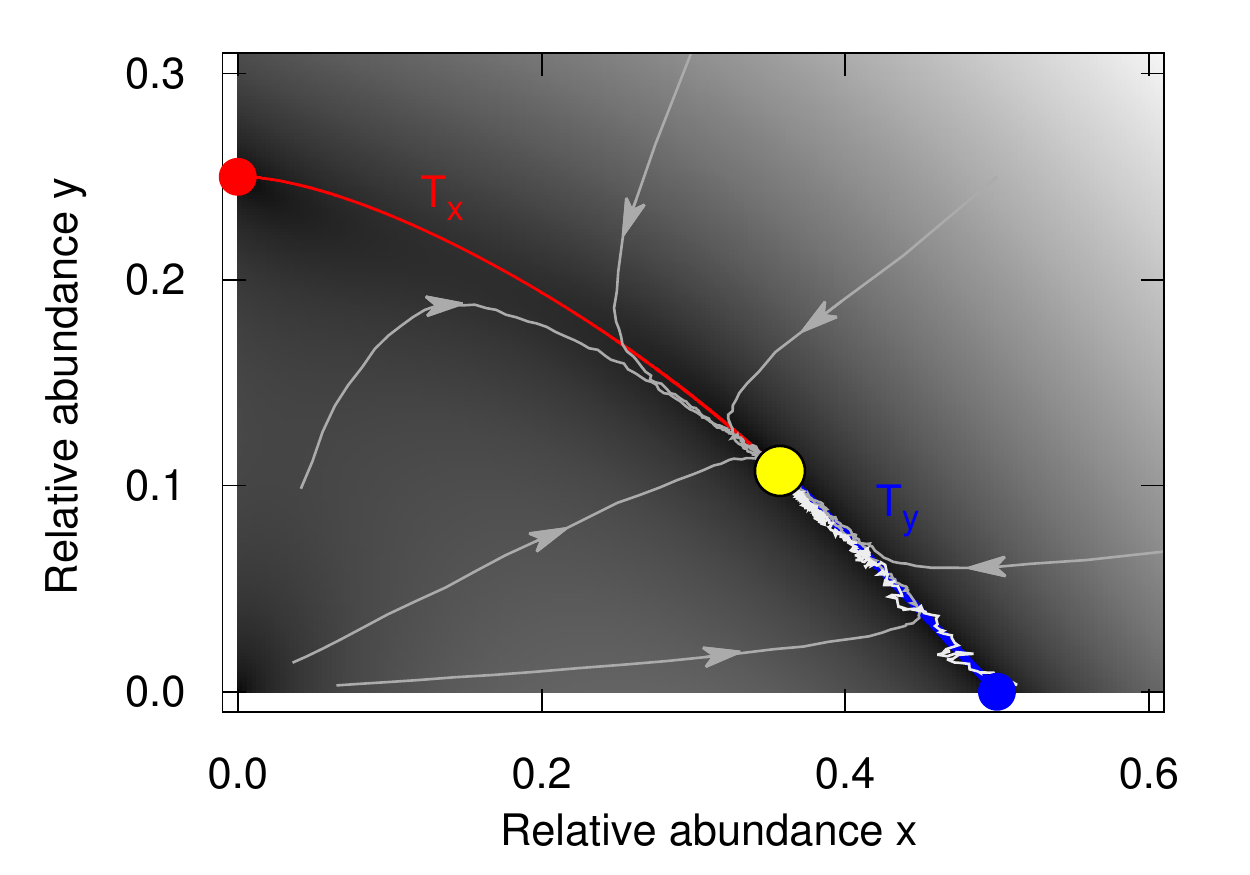}
\vspace{-0.5cm}
\caption{(Color online)
The most probable trajectories to extinction in $x-y$ space.
Trajectories are projections from the 4-dimensional phase space on the abundance space.
Red and blue circles represent extinction points for type $X$ and $Y$, respectively.
We compute $\mathcal{S}_i$ along the trajectories from the coexistence (yellow) to extinction and find that $Y$ is more likely to go extinct first ($\mathcal{S}_x \approx 0.032$, and $\mathcal{S}_y \approx 0.006$).
To confirm our result, we simulate the surviving-averaged extinction trajectory (10000 realizations) shown as a a bright line.
We only use realizations whereby extinction occurs within $t=1000$ for $M=400$.
The simulation result matches the WKB result very well
(parameters $a=1, b=0.75, c=1.25$, $d=0.5$, $\lambda_b=0.9$, and $\lambda_d=0.4$).
}
\label{f.path}
\end{figure}
Because the first extinction time $\tau$ is exponentially large in $M$~\cite{gottesman:PRE:2012,gabel:PRE:2013}, 
we can set $1/\tau$ to zero for large $M$.
Therefore, the most probable extinction trajectories are captured by the effective Hamiltonian with zero energy.
From the derivatives of effective Hamiltonian, we derive the equations of motion in the phase space \cite{goldstein:book:2002}, 
$\dot x= \partial_{p_x} H$, 
$\dot y= \partial_{p_y} H$, 
$\dot p_x= - \partial_{x} H$, and
$\dot p_y= - \partial_{y} H$, 
	\begin{equation}
	\begin{split}
	\dot{x}=&x(\lambda_be^{p_x}-\lambda_de^{-p_x})-\tfrac{x^2}{a}e^{-p_x}-\tfrac{xy}{b}e^{-p_x},  \\
	\dot{y}=&y(\lambda_be^{p_y}-\lambda_de^{-p_y})-\tfrac{y^2}{d}e^{-p_y}-\tfrac{xy}{c}e^{-p_y}, \\
	\dot{p_x}=&\lambda_b(1\!-\!e^{p_x})\!+\!\lambda_d(1\!-\!e^{\!-p_x})
		\\&{} \!+\!\tfrac{2x(1\!-\!e^{\!-p_x})}{a}\!+\!\tfrac{y}{b}(1\!-\!e^{\!-p_x})\!+\!\tfrac{y}{c}(1\!-\!e^{\!-p_y}), \\
	\dot{p_y}=&\lambda_b(1\!-\!e^{p_y})\!+\!\lambda_d(1\!-\!e^{\!-p_y})
		\\&{} \!+\!\tfrac{2y(1\!-\!e^{\!-p_y})}{d}\!+\!\tfrac{x}{b}(1\!-\!e^{\!-p_x})\!+\!\tfrac{x}{c}(1\!-\!e^{\!-p_y}).	
	\end{split}\raisetag{4\baselineskip}
	\label{e.eom}
	\end{equation}
For $p_x\!=\!p_y\!=\!0$, deterministic equations are recovered.
Since we are interested in the trajectory to extinction from the coexistence quasi-steady state, 
the system initially starts from the coexistence with 
$(x, y, p_x, p_y)=(\frac{\lambda ac[d-b]}{ad-bc},\frac{\lambda bd[a-c]}{ad-bc},0,0)$.
There are eight fixed points of Eq.~(\ref{e.eom}) related to extinction states ($x=0$ or $y=0$) with zero-energy.
Three of these points describe deterministic trajectories and are thus of no further interest here.
Two of the points describe single populations. 
One fixed point describes extinction of both species almost at the same time, which occurs with negligible probability.
Hence, we focus on the two extinction fixed points,
$(a\lambda, 0, 0, \ln{[\frac{a\lambda+c\lambda_d}{c\lambda_b}]})$ and $(0, d\lambda, \ln{[\frac{d\lambda+b\lambda_d}{b\lambda_b}]} , 0)$.
We will find extinction trajectories from the coexistence to each extinction point.

We numerically find trajectories to extinction using the Chernykh-Stepanov numerical iteration
algorithm~\cite{chernykh:PRE:2001,elgart:PRE:2004, lohmar:PRE:2011, lohmar:jpam:2017}: 
Coordinates and momenta are changed in turn. 
Coordinates are adjusted forward in time while momenta are adjusted backward.
This procedure is iterated until the trajectory no longer changes.
To do that, we first set all coordinates to the coexistence coordinates
 [$x(t)=x(0)$ and $y(t)=y(0)$ for all $t$] while momenta are set to the final values of the extinction point.
Note that we need a long time sequence to capture extinction trajectories~\cite{elgart:PRE:2004}.
After setting the values, momenta are updated using the equations of motion backward in time for fixed coordinates. 
Using this updated momenta, coordinates are updated forward in time.
As momenta may diverge during numerical integration, we update each momentum in turn.  
After many iterations, the trajectory remains unchanged.

To address which type is more likely to go extinct first, we compute the transition rates from the coexistence to the single-type populations.  
As $M$ increases, the effective potential becomes steeper and extinction takes longer.
As a result, almost every extinction occurs along the most probable trajectories for large $M$.
Hence, the extinction rates $\mathcal{R}_x$ and $\mathcal{R}_y$ of species $X$ and $Y$ can be calculated from
	\begin{equation}
		\mathcal{R}_x\propto \exp(-MS[\mathcal{T}_x]),~ 		\mathcal{R}_y\propto \exp(-MS[\mathcal{T}_y])
	\label{e.r},
	\end{equation}	
where $S[\mathcal{T}]$ is an integral along the extinction trajectory,
	\begin{equation}
	\begin{split}
	S[\mathcal{T}]= &\int_{0}^{\infty} dt(p_x\dot{x} + p_y\dot{y})
	\end{split}.
	\label{e.s}
	\end{equation}
On the trajectory $\mathcal{T}_x$, $X$ goes to extinction first,
while on $\mathcal{T}_y$, $Y$ goes to extinction first.
For large $M$, the exponential term dominates the pre-factor in Eq.~(\ref{e.r}), and thus the most probable trajectory is determined by the minimum $S[\mathcal{T}_i]$.  

We show the most probable trajectories to extinction in Fig.~\ref{f.path} at given parameters. 
The most probable extinction trajectories are close to the paths which minimize the potential gradient, 
but not identical (see Fig.~\ref{f.path}).
For the respective parameter set, $\mathcal{S}_x$ is larger than $\mathcal{S}_y$, 
where $\mathcal{S}_i = S[\mathcal{T}_i]$,
and thus the extinction mostly occurs along the trajectory $\mathcal{T}_y$:
$Y$ goes extinct first, and we obtain the quasi-steady state of the single $X$ population.
Eventually, also $X$ goes extinct \cite{elgart:PRE:2004}. 
We also obtain the average extinction trajectory from many realizations of the 
stochastic process. 
The most probable trajectory matches the simulation result very well (see Fig.~\ref{f.path}).

Even though $\mathcal{S}$ is not linear in the trajectory length, our results imply that distances from the coexistence to the the extinction points of $X$ and $Y$,  $l_x$ and $l_y$, may affect which type goes extinct first.
To find which factor is more crucial for determining the first extinction type, we calculate $f=\mathcal{S}_x/\mathcal{S}_y$ for various parameters.
If $f$ is larger than unity, the trajectory $\mathcal{T}_y$ is more likely to happen than $\mathcal{T}_x$.

\begin{figure}[t]
\includegraphics[width=.5\textwidth]{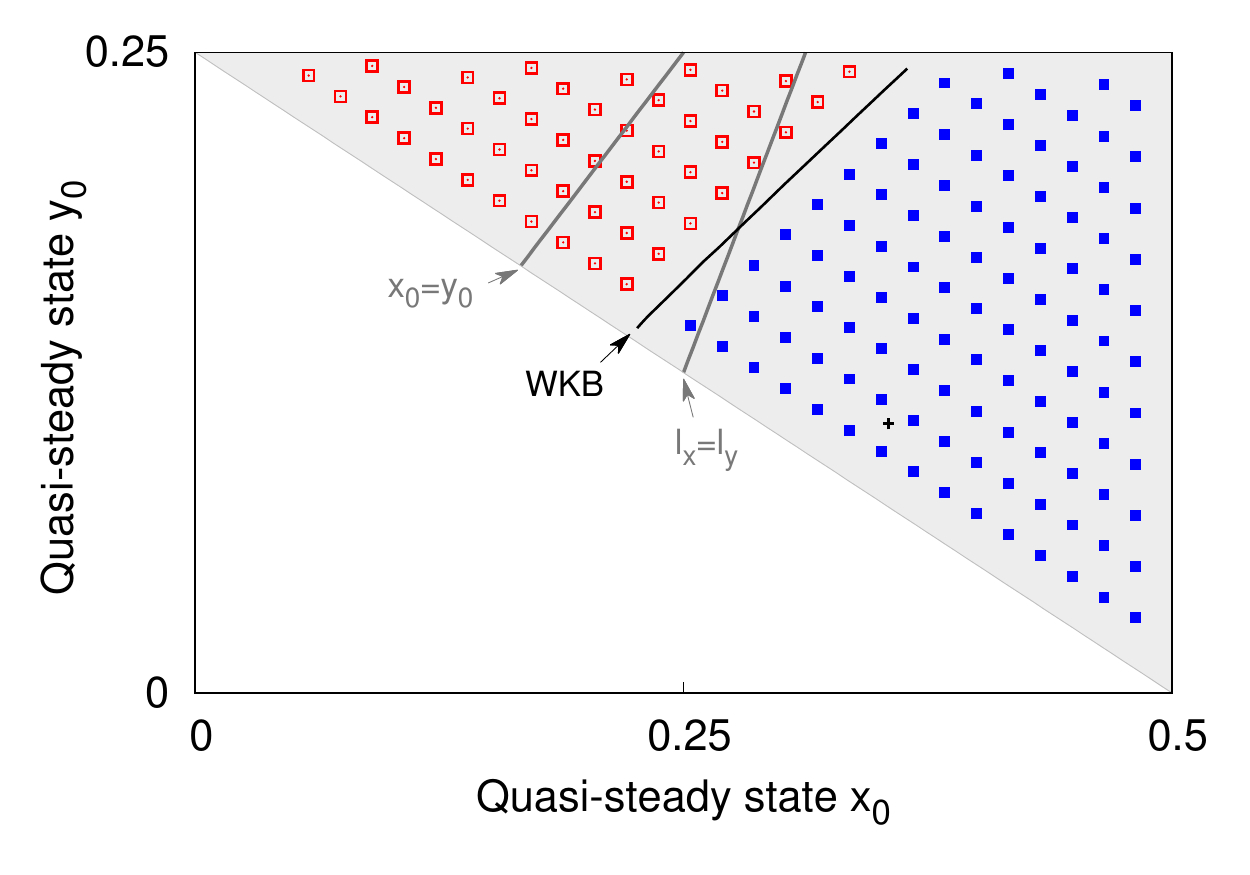}
\vspace{-0.7cm}
\caption{(Color online)
Possible coexistence state for $a=1$ and $d=0.5$ with fixed $\lambda=0.5$ (upper triangle region). 
Three indicators, $x_0=y_0$, $l_x=l_y$, and $f=1$ (WKB), are represented. 
Stochastic simulation results are shown as symbols.
Empty (filled) squares indicate the extinction of $X$ $(Y)$.
There is a region where type $X$ is more abundant, but is more likely to go extinct first.
More interestingly, although the abundances and distance indicator predict the first extinction of $Y$, there is a region where $X$ is more likely to go extinct first.
While distance can be a better indicator for extinction than abundances, only the WKB method correctly predicts the first extinction type.
}
\label{f.para}
\end{figure}

The coexistence state $(x_0, y_0)$ is determined by payoffs, and thus possible $x_0$ and $y_0$ are restricted.
Figure~\ref{f.para} shows possible $x_0$ and $y_0$ 
and the separation line ($f=1$) where both types go extinct at the same rate. 
This shows that, as a rule of thumb, the distance from the extinction point is a better predictor
of extinction probabilities than the abundance in equilibrium. 
More importantly, however, the path to extinction is not determined by these factors --- instead, it 
depends on the trajectory from quasi-stationary coexistence to extinction with zero-energy.  
A compelling examples are the parameters in Fig.~\ref{f.para}
where extinction is, maybe counterintuitively, most likely of the more abundant type which is further away from the extinction state.

We consider a stochastic model where pairwise interactions are reflected in death rates.
For coexistence games, two types coexist in populations for a long time.
Due to stochasticity, however, extinctions always occur after a sufficiently long time.
Our focus is the most probable trajectory to extinction from the coexistence of two types.
By mapping our reaction system to the effective Hamiltonian system using the WKB method,
we extract the rare event information, and get the most probable trajectory to extinction.
Mainly, we analyse which type is more likely to go extinct first between two types.
Because of the pathway to extinction, there is tendency that the type closer to its quasi-steady state of the single-type population 
is more likely to go to extinction first.
However, there is a region where distance fails to predict the first extinction type --- 
only the WKB method makes a correct prediction in this case.

We apply a Hamiltonian framework to evolutionary game dynamics.
Although reaction systems have already been used for describing biological populations~\cite{murray:book:2007,gabel:PRE:2013} and ecological systems~\cite{okubo:book:1980}, our model can be directly applied within stochastic evolutionary game dynamics, leading to results that are out of reach without this approach.

We thank Alex Kamenev and Weini Huang for fruitful discussions.
\bibliographystyle{apsrev4-1}
%

\end{document}